\documentclass[11pt]{cernrep}
\usepackage{graphicx}
\usepackage{here}
\def\lsim{~\raise0.3ex\hbox{$<$\kern-0.75em\raise-1.1ex\hbox{$\sim$}}~}
\def\gsim{~\raise0.3ex\hbox{$>$\kern-0.75em\raise-1.1ex\hbox{$\sim$}}~}

\begin{document}

\title{Global DGLAP fit analyses of the nPDF: EKS98 and HKM\footnote{~~Contribution to CERN Yellow Report on Hard Probes in Heavy Ion Collisions at the LHC.}}

\author{K.J. Eskola$^{a,b}$, H. Honkanen$^{a,b}$, V.J. Kolhinen$^a$, 
C.A. Salgado$^c$}

\institute{\em{
$^a$Department of Physics, P.O. Box 35, FIN-40014 University of Jyv\"askyl\"a, Finland\\
$^b$Helsinki Institute of Physics, P.O. Box 64, FIN-00014 University of Helsinki, Finland, \\
$^c$CERN, Theory Division, CH-1211 Geneva, Switzerland
}}

\maketitle

\vspace{-4.8cm}
\begin{flushright}
  HIP-2003-09/TH
\end{flushright}
\vspace{3.3cm}

\begin{abstract}
The DGLAP analyses of the nuclear parton distribution functions (nPDF)
based on the global fits to the data are reviewed, and the results
from EKS98 and HKM are compared. The usefulness of measuring hard
probes in $pA$ collisions, at the LHC in particular, is demonstrated.
\end{abstract}

\section{Introduction}

Inclusive cross sections for hard processes $A+B\rightarrow c+X$
involving a sufficiently large scale $Q\gg\Lambda_{\rm QCD}$ are
computable by using collinear factorization. In the leading twist
approximation power corrections $\sim 1/Q^2$ are neglected and
\begin{eqnarray}
& \displaystyle 
   d\sigma(Q^2,\sqrt s)_{AB\rightarrow c+X} =
   \sum_{i,j=q,\bar q,g} 
   \bigg[ Z_Af_i^{p/A}(x_1,Q^2)+(A-Z_A)f_i^{n/A}(x_1,Q^2)\bigg] 
   \otimes \nonumber  \\
 & \otimes\bigg[ Z_Bf_j^{p/B}(x_2,Q^2)+(B-Z_B)f_j^{n/B}(x_2,Q^2)\bigg]
   \otimes d\hat \sigma(Q^2,x_1,x_2)_{ij\rightarrow c+x}
\label{hardAA}
\end{eqnarray}
where $A,B$ are the colliding hadrons or nuclei containing $Z_A$ and
$Z_B$ protons correspondingly, $c$ the produced parton, $x$ and $X$
anything, $d\hat \sigma(Q^2,x_1,x_2)_{ij\rightarrow c+x}$ the
perturbatively calculable differential cross section for the
production of $c$ at the scale $Q$, and $x_{1,2}\sim Q/\sqrt s$ the
fractional momenta of the colliding partons $i$ and $j$. The number
distribution function of the parton flavour $i$ of the protons
(neutrons) in $A$ is denoted as $f_i^{p/A}$ $(f_i^{n/A})$, and
similarly for partons $j$ in $B$.

In the leading twist approximation multiple scattering of the bound
nucleons does occur but all collisions are independent, correlations
between partons from the same object $A$ are neglected, and only
one-parton densities are needed. The parton distribution functions
(PDF) $f_i^{p/A}$ are universal quantities and applicable in all
collinearly factorizable processes. The PDF cannot be computed by
perturbative methods, and so far it has not been possible to compute
them from first principles, either. Thus, nonperturbative input from
data on various hard processes is needed for the extraction of the
PDF.  However, once the PDF are known at some initial (lowest) scale
$Q_0\gg\Lambda_{\rm QCD}$, the QCD perturbation theory predicts the
scale evolution of the PDF to other (higher) values of $Q^2$ in form
of the Dokshitzer-Gribov-Lipatov-Altarelli-Parisi (DGLAP) equations
\cite{Dokshitzer:sg}.

The method to extract the PDF from experimental data is well
established in the case of the free proton: the initial
(non-perturbative) distributions are parametrized at some $Q_0^2$, and
evolved to higher scales according to the DGLAP equations. Comparison
with the data is made at various regions of the $(x,Q^2)$-plane, and
the parameters of the initial distributions $f_i^{p}(x,Q_0^2)$ become
fixed when the best global fit is found. The data from deeply
inelastic $lp$ scattering (DIS) are of main importance in these global
DGLAP fits, especially the HERA data at small values of $x$ and
$Q^2$. The sum rules for momentum, charge and baryon number give
further constraints. In this way, through the global DGLAP fits,
groups like MRST \cite{Martin:2001es}, CTEQ \cite{Pumplin:2002vw} or
GRV \cite{Gluck:1998xa} obtain their well-known sets of the PDF of the
free proton.

The nuclear parton distribution functions (nPDF) differ in magnitude
from the PDF of the free proton. In the measurements of the structure
function $F_2^A = Z_AF_2^{p/A}+ (A-Z_A)F_2^{n/A}$ of nuclear targets
in DIS (see e.g. \cite{Arneodo:1992wf} for references) and especially
of the ratio
\begin{equation}
R_{F_2}^A(x,Q^2) \equiv
\frac{\frac{1}{A}{d\sigma^{lA}}/{dQ^2dx}}{\frac{1}{2}{d\sigma^{lD}}/{dQ^2dx}}
\approx \frac{\frac{1}{A}F_2^A(x,Q^2)}{\frac{1}{2}F_2^D(x,Q^2)},
\end{equation}
the following nuclear effects have been discovered as a function of 
Bjorken-$x$:
\begin{itemize}
\item shadowing; a depletion $R_{F_2}^A<1$  at $x \lsim 0.1$,
\item anti-shadowing; an excess $R_{F_2}^A>1$ at $0.1 \lsim x \lsim 0.3$,
\item EMC effect; a depletion at $0.3 \lsim x\lsim0.7$, and
\item Fermi motion; an excess towards $x\rightarrow1$ and beyond.
\end{itemize}
The $Q^2$ dependence of $R_{F_2}^A$ is weaker and has thus been more
difficult to measure. Data with high enough precision, however, exist:
NMC has some years ago discovered a clear $Q^2$ dependence in the
ratio $d\sigma^{\mu{\rm Sn}}/d\sigma^{\mu{\rm C}}$
\cite{Arneodo:1996ru}, i.e. the scale dependence of the ratio
$F_2^{\rm Sn}/F_2^{\rm C}$, at $x\gsim 0.01$. Since $F_2^{p(n)/A}=
\sum_{q} e_q^2x[f_q^{p(n)/A}+f_{\bar q}^{p(n)/A}]+ {\cal
O}(\alpha_s)$, the nuclear effects in the ratio $R_{F_2}^A$ directly
translate into nuclear effects in the parton distributions;
$f_i^{p/A}\ne f_i^{p}$.

\begin{figure}[htb]
\begin{center}
\vspace{-3cm}
\includegraphics[width=12cm]{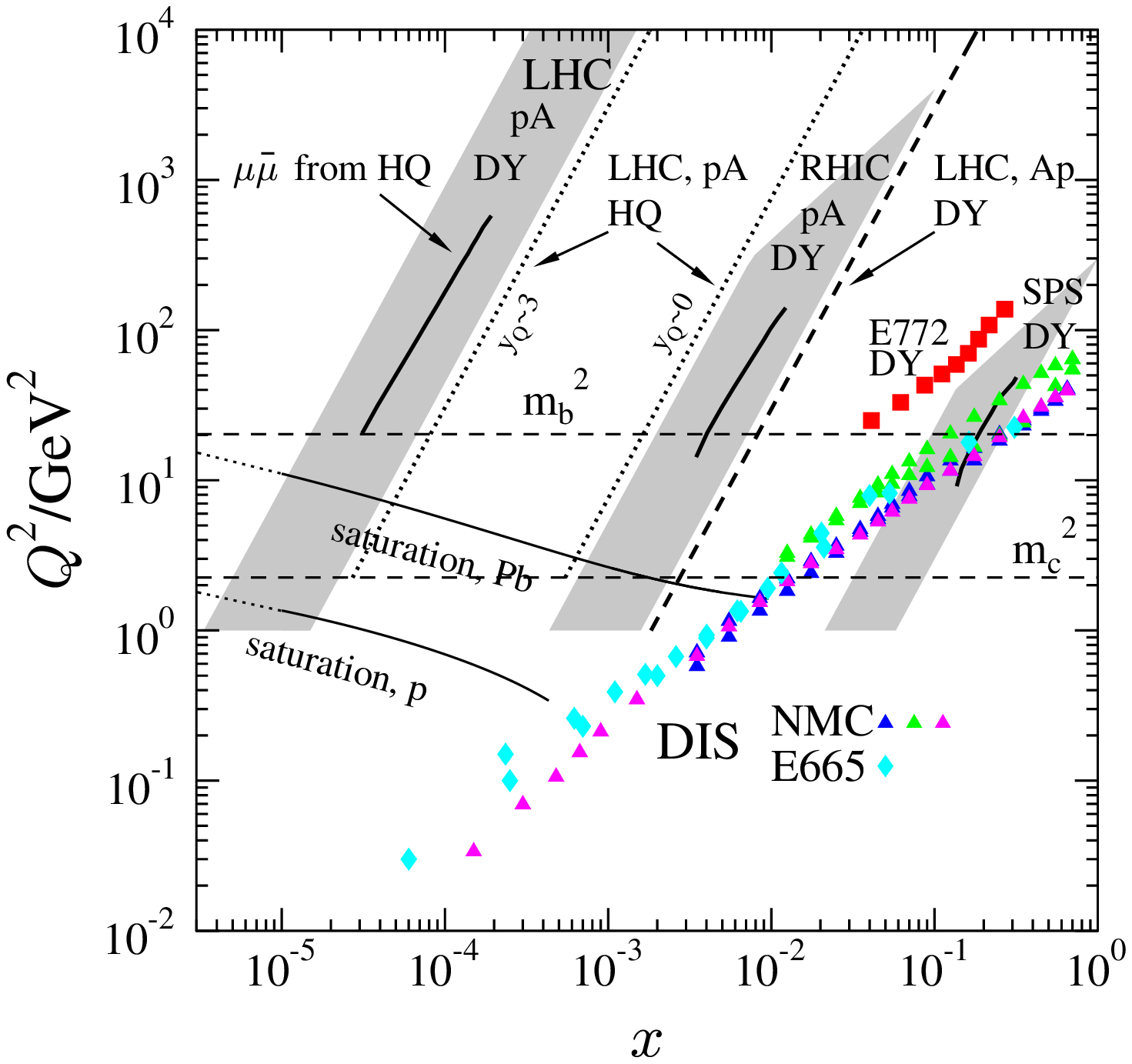} 
\vspace{-3.5cm}
\caption{The average values of $x$ and $Q^2$ of the DIS data from NMC
\cite{Amaudruz:1995tq,Arneodo:1995cs,Arneodo:1996rv} (triangles) and
E665 \cite{Adams:1992nf,Adams:1995is} (diamonds) in $lA$, and of $x_2$
and $M^2$ of the DY dilepton data \cite{Alde:im} (squares) in
$pA$. The heavy quark mass scales are shown by the horizontal dashed
lines.  The initial scale $Q_0^2$ is $m_c^2$ in EKRS and 1 GeV$^2$ in
HKM. The rest of the figure: see the text in
Sec.~\ref{future}.  }
\label{fig:q2_vs_x}
\end{center}
\end{figure}

The nPDF $f_i^{p/A}$ also obey the DGLAP equations in the large-$Q^2$
limit, and they can be determined by using a similar global DGLAP fit
procedure as in the case of the PDF of the free proton.  Pioneering
studies of the DGLAP evolution of the nPDF are
e.g. \cite{Qiu:wh,Frankfurt:xz,Eskola:1992zb,Kumano:1994pn}. References
for various other studies of perturbative evolution of the nPDF and
also to simpler $Q^2$-independent parametrizations of the nuclear
effects in the PDF can be found e.g. in
\cite{Eskola:2002us,Eskola:2001ms}. The nuclear case is, however, more
complicated because of additional variables, the mass number $A$ and
the charge $Z$, and, because the number of data points available in
the perturbative region is more limited than for the PDF of the free
proton.  The DIS data plays the dominant role in the nuclear case as
well. However, as illustrated by Fig.~\ref{fig:q2_vs_x}, no data are
available from nuclear DIS experiments below $x\lsim 5\cdot 10^{-3}$
at $Q^2\gsim1$~GeV$^2$. This makes the determination of the nuclear
gluon distributions especially difficult. Further constraints for the
global DGLAP fits of the nPDF can be obtained from e.g. the Drell-Yan
(DY) process measured in fixed-target $pA$ collisions
\cite{Alde:im,Vasilev:1999fa}. Currently, there are two sets of nPDF
available which are based on the global DGLAP fits to the data: (i)
EKS98 \cite{Eskola:1998df,Eskola:1998iy} (the code in
\cite{EKS98_code,Plothow-Besch:ci}), and (ii) HKM \cite{Hirai:2001np}
(the code in \cite{HKM_code}). We shall compare the main features of
these two analyses and comment on their differences below.

\section{Comparison of EKS98 and HKM}

\subsection{EKS98 and overview of constraints available from data}

The parametrization EKS98 prepared in \cite{Eskola:1998df} is based on
the results of the DGLAP analysis in \cite{Eskola:1998iy} and its
follow-up in \cite{Eskola:1998df}. We shall here refer to these
together as EKRS.  In the EKRS approach the nPDF, the parton
distributions of the bound {\em protons}, $f_i^{p/A}$ , are defined
through the modifications of the corresponding distributions in the
free proton,
\begin{equation}
R_i^A(x,Q^2)\equiv {f_i^{p/A}(x,Q^2)\over f_i^p(x,Q^2)},
\label{eqratios}
\end{equation}
where the PDF of the free proton are from a known set, such as CTEQ,
MRS or GRV. As in the case of the free nucleons, for isoscalar nuclei
the parton distributions of bound neutrons are obtained through
isospin symmetry, $f_{u(\bar u)}^{n/A}=f_{d(\bar d)}^{p/A}$ and
$f_{d(\bar d)}^{n/A}=f_{u(\bar u)}^{p/A}$.  Although exact only for
isoscalar and mirror nuclei, this is expected to be a good first
approximation for all nuclei.

To simplify the determination of the input nuclear effects for valence
and sea quarks, the following flavour-independent {\em initial}
modifications are assumed: $R_{u_V}^A(x,Q_0^2)\approx
R_{d_V}^A(x,Q_0^2) \approx R_V^A(x,Q_0^2)$, and $R_{\bar
u}^A(x,Q_0^2)\approx R_{\bar d}^A(x,Q_0^2) \approx
R_{s}^A(x,Q_0^2)\approx R_S(x,Q_0^2)$.  Thus only three independent
initial ratios, $R_V^A$, $R_S^A$ and $R_G^A$ are to be determined at
$Q_0^2=m_c^2=2.25$~GeV$^2$. Note also that the approximations above
are needed and used {\em only} at $Q_0^2$ in \cite{Eskola:1998iy}. In
the EKS98-parametrization \cite{Eskola:1998df} of the DGLAP-evolved
results, it was observed that to a good approximation $R_{u_V}^A =
R_{d_V}^A$ and $R_{\bar u}^A=R_{\bar d}^A$ for all $Q^2$.  Further
details of the EKRS analysis can be found in \cite{Eskola:1998iy},
here we summarize the constraints available in each region of
$x$. Consider first quarks and antiquarks:

\begin{itemize}
\item 
At { $x\gsim 0.3$} the DIS data constrains only the ratio $R_V^A$:
valence quarks dominate $F_2^A$, so $R_{F_2}^A\approx R_V^A$ but
$R_S^A$ and $R_G^A$ are left practically unconstrained. An EMC effect
is, however, assumed also for the initial $R_S^A$ and $R_G^A$ since in
the scale evolution the EMC effect of valence quarks is transmitted to
gluons and further on to sea quarks \cite{Eskola:1992zb}. In this way the
nuclear modifications $R_i^A$ also remain stable against the
evolution. 

\item 
At {$0.04\lsim x \lsim 0.3$} both the DIS and DY data constrain
$R_S^A$ and $R_V^A$ but from different regions of $Q^2$, as shown in
Fig. \ref{fig:q2_vs_x}. In addition, conservation of baryon number
restricts $R_V^A$. The use of DY data \cite{Alde:im} is essential in
order to fix the relative magnitude of $R_V^A$ and $R_S^A$, since the
DIS data alone cannot distinguish between them. As a result, no
antishadowing appears in $R_S^A(x,Q_0^2)$.

\item 
At { $5\cdot10^{-3}\lsim x\lsim0.04$} only DIS data exist in the
region $Q\gsim1$ GeV where the DGLAP analysis can be expected to
apply. Once $R_V^A$ is fixed by the DIS and DY data at larger $x$, the
magnitude of nuclear valence quark shadowing in the EKRS approach
follows from the conservation of baryon number. As a result of these
contraints, nuclear valence quarks in EKRS are less shadowed than the
sea quarks, $R_V^A>R_S^A$.

\item 
At $x\lsim 5\cdot10^{-3}$, as shown by Fig. \ref{fig:q2_vs_x}, the DIS
data for the ratio $R_{F_2}^A$ lie in the region $Q\lsim1$ GeV where
the DGLAP equations are unlikely to be applicable in the nuclear
case. Indirect constraints, however, {\em can} be obtained by
noticing: (i) At the smallest values of $x$ (where $Q\ll1$~GeV),
$R_{F_2}^A$ depends only very weakly, if at all, on $x$
\cite{Arneodo:1995cs,Adams:1992nf}. (ii) $\partial (F_2^{\rm
Sn}/F_2^{\rm C})/\partial \log Q^2>0$ at $x\sim0.01$
\cite{Arneodo:1996ru}, indicating that $\partial R_{F_2^A}/\partial
\log Q^2>0$. (iii) Negative $\log Q^2$-slopes of $R_{F_2}^A$ at
$x\lsim 5\cdot10^{-3}$ have not been observed \cite{Arneodo:1995cs}.
Based on these experimental facts, it is assumed in the EKRS approach
that the sign of the $\log Q^2$-slope of $R_{F_2}^A$ remains
non-negative at $x<0.01$ and therefore the DIS data
\cite{Arneodo:1995cs,Adams:1992nf} in the non-perturbative region
gives a {\em lower bound} for $R_{F_2}^A$ at $Q_0^2$ at very small
$x$.  The sea quarks dominate over the valence quarks at small values
of $x$, so only $R_S^A$ becomes constrained by the DIS data; $R_V^A$
remains restricted by baryon number conservation.

\end{itemize}

\noindent 
Pinning down the nuclear gluon distributions is difficult in the
absense of stringent direct constraints. What is available in
different regions of $x$ can be summarized as follows:

\begin{itemize}

\item 
At $x\gsim0.2$ no experimental constraints are currently available for
the gluons. Conservation of momentum is used as an overall constraint
but this alone is not sufficient to determine whether an EMC effect
exists for gluons or not: only about 30 \% of the gluon momentum comes
from $x\gsim0.2$, so fairly sizable changes in $R_G^A(x,Q_0^2)$ in
this region can be compensated by smaller changes at $x<0.2$ without
violating the constraints discussed below.  As mentioned above, in the
EKRS approach an EMC effect is initially assumed for $R_G$. This
guarantees a stable scale evolution; the EMC effect remains there also
at larger $Q^2$.

\item 
At $0.02 \lsim x \lsim0.2$ the $Q^2$ dependence of the ratio $F_2^{\rm
Sn}/F_2^{\rm C}$ measured by NMC \cite{Arneodo:1996ru} sets the
currently most important constraint for $R_G^A$, as first pointed out
in \cite{Gousset:1996xt}. In the small-$x$ region where gluons
dominate the DGLAP evolution, the $Q^2$ dependence of $F_2(x,Q^2)$ is
dictated by the gluon distribution as \cite{Prytz:1993vr} $\partial
F_2(x,Q^2)/\partial \log Q^2\sim \alpha_s xg(2x,Q^2)$. This leads to
\cite{Eskola:1998iy,Eskola:2002us} $\partial R_{F_2}^A(x,Q^2)/\partial
\log Q^2 \sim \alpha_s [R_G^A(2x,Q^2)-R_{F_2}^A(x,Q^2)]
xg(2x,Q^2)/F_2^D(x,Q^2)$.  The $\log Q^2$ slopes of $F_2^{\rm
Sn}/F_2^{\rm C}$ measured by NMC \cite{Arneodo:1996ru} therefore
constrain $R_G^A$. Especially, as discussed in detail in
\cite{Eskola:1998iy,Eskola:2002us}, the positive $\log Q^2$-slope of
$F_2^{\rm Sn}/F_2^{\rm C}$ measured by NMC indicates that
$R_G^A(2x,Q_0^2)>R_{F_2}^A(x,Q_0^2)$ at $x\sim0.01$. Thus, within the
DGLAP framework, a gluon shadowing at $x\sim 0.01$ which would be much
stronger than the shadowing of antiquarks, such as suggested e.g.  in
\cite{Li:2001xa}, is not supported by the NMC data. The antishadowing
in the EKRS gluons follows from the constraint
$R_G^A(0.03,Q_0^2)\approx 1$ imposed by the NMC data (see also
\cite{Gousset:1996xt}) combined with the requirement of momentum
conservation. The EKRS antishadowing is consistent with the E789 data
\cite{Leitch:1994vc} on $D$-meson production in $pA$ collisions
(notice, however, the large error bar of the data point), and it seems
to be supported by the $J/\Psi$ production in DIS, measured by NMC
\cite{Amaudruz:1991sr}.

\item 
At $x\lsim0.02$, stringent experimental constraints do not exist for
the nuclear gluons at the moment. It should be emphasized, however,
that the initial $R_G^A$ in this region is directly connected with the
initial $R_{F_2}^A$. As discussed above, related to quarks at small
$x$, taking the DIS data on $R_{F_2}^A$ in the non-perturbative region
as a lower limit for $R_{F_2}^A$ at $Q_0^2$ corresponds to $\partial
R_{F_2^A}/\partial \log Q^2\ge0$, and thus $R_G^A(x<0.02,Q_0^2)\ge
R_{F_2}^A(x/2,Q_0^2)\ge R_{F_2}^A(x/2,Q^2\ll1\,{\rm GeV}^2)$. The
observation in \cite{Eskola:1998iy} was that setting
$R_G^A(x,Q_0^2)\approx R_{F_2}^A(x,Q_0^2)$ at $x\lsim 0.01$ fulfills
this constraint. This approximation remains fairly good even after the
DGLAP evolution from $Q_0\sim 1$ GeV to $Q\sim100$ GeV, see
\cite{Eskola:1998df}.

\end{itemize}

\bigskip

As explained in detail in \cite{Eskola:1998iy}, in the EKRS approach
the initial ratios $R_V^A(x,Q_0^2)$, $R_S^A(x,Q_0^2)$ and
$R_G^A(x,Q_0^2)$ are constructed piecewize in different regions of
$x$. Initial nPDF are computed at $Q_0^2$, LO DGLAP evolution to
higher scales is performed and comparison with DIS and DY data is
made. The parameters in the input ratios are iteratively changed until
the best global fit to the data is achieved. The determination of the
input parameters in EKRS has so far been done only manually, the best
overall fit is determined by eye. For the quality of the obtained fit,
see the detailed comparison with the data in Figs. 4-10 of
\cite{Eskola:1998iy}.  The parametrization EKS98 \cite{Eskola:1998df}
of the nuclear modifications $R_i^A(x,Q^2)$ was prepared on the basis
of the results in \cite{Eskola:1998iy}. It was also shown that when
the PDF of the free proton where changed from GRVLO
\cite{Gluck:1991ng} to CTEQ4L \cite{Lai:1996mg} (differing from each
other considerably), the changes induced to $R_i^A(x,Q^2)$ were within
a few percents \cite{Eskola:1998df}.  Therefore, accepting this range
of uncertainty, the EKS98 parametrization can be used together with
any (LO) set of the PDF of the free proton.

\subsection{The HKM analysis}

In principle, the definition of the nPDF in the HKM analysis
\cite{Hirai:2001np} differs slightly from that in EKRS: instead of the
PDF of the bound protons, HKM define the nPDF as the {\em average}
distributions of each flavour $i$ in a nucleus $A$: $f_i^A(x,Q^2) =
(Z/A)f_i^{p/A}(x,Q^2) + (1-Z/A)f_i^{n/A}(x,Q^2)$. Correspondingly, the
HKM nuclear modifications at the initial scale $Q_0^2=1$~GeV$^2$ are
then defined through
\begin{equation}
f_i^A(x,Q_0^2) = 
w_i(x,A,Z)[(Z/A)f_i^{p}(x,Q_0^2) + (1-Z/A)f_i^{n}(x,Q_0^2)].
\end{equation}
In practice, however, since in the EKS98 parametrization one sets
$R_{u_V}^A = R_{d_V}^A$ and $R_{\bar u}^A=R_{\bar d}^A$, also the
EKS98 modifications represent average modifications. Flavour-symmetric
sea quark distributions are assumed in HKM, whereas the flavour
asymmetry of the sea quarks in EKRS follows from that of the free
proton.

An improvement relative to EKRS is that the HKM method to extract the
initial modifications $w_i(x,A,Z)$ at $Q_0^2$ is more automatic and
more quantitative in the statistical analysis; the HKM analysis is
strictly a minimum-$\chi^2$ fit. Also, with certain assumptions of a
suitable form (see \cite{Hirai:2001np}) for the initial modifications
$w_i$, the number of parameters has been brought conveniently down to
seven. The form used is
\begin{equation}
w_i(x,A,Z)=1+\left(1-{1\over A^{1/3}}\right){a_i(A,Z)+H_i(x)\over
(1-x)^{\beta_i}},
\label{eqSalgado2}
\end{equation}
where $H_i(x)=b_ix+c_ix^2$ (analysis with a cubic polynomial is also
performed, with similar results). Due to the flavour-symmetry, the sea
quark parameters are identical for all flavours. For valence quarks,
conservation of charge $Z$ (not used in EKRS) and baryon number $A$
are required; this fixes $a_{u_V}$ and $a_{d_V}$. In the case of
non-isoscalar nuclei $w_{u_V}\ne w_{d_V}$. Also momentum conservation
is imposed; this fixes $a_g$. Taking $\beta_V$, $b_V$, $c_V$ to be the
same for $u_V$ and $d_V$, and $\beta_{\bar q}=\beta_g=1$, $b_g=-2c_g$,
the remaining seven parameters $b_V$, $c_V$, $\beta_V$, $a_{\bar q}$
$b_{\bar q}$, $c_{\bar q}$, $c_g$ are determined by a global DGLAP fit
which minimizes the $\chi^2$.

\subsection{The comparison}

Irrespective of whether the best fit is found automatically or by eye,
the basic procedure to determine the nPDF in the EKRS and HKM analyses
is the same.  The results obtained for the nuclear effects are,
however, quite different, as can be seen in the comparison at $Q^2 =
2.25$~GeV$^2$ shown in Fig. \ref{fig:initial}. The main reason for
this is that different data sets are used:

\begin{figure}[htb]
\begin{center}
\vspace{0cm}
\includegraphics[width=13cm]{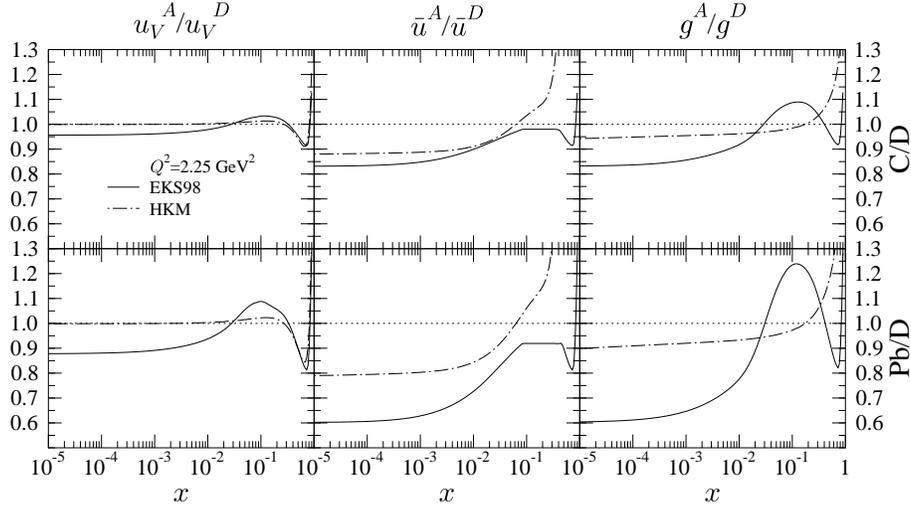} 
\vspace{0cm}
\caption{The nuclear modifications for the $u_V$, $\bar u$ and gluon
distributions in EKS98 (solid) and in HKM (dotted dashed) at $Q^2 =
2.25$~GeV$^2$. Upper panels: C/D. Lower panels: Pb/D, with isoscalar
Pb.  }
\label{fig:initial}
\end{center}
\end{figure}

\begin{itemize}

\item The HKM analysis \cite{Hirai:2001np} uses only the DIS data,
whereas EKRS include also the DY data from $pA$ collisions. As
explained above, the DY data is very important in the EKRS analysis in
fixing the relative modifications of valence and sea quarks at
intermediate $x$. Preliminary results reported in \cite{Kumano:2002ii}
show that when the DY data is included in the HKM analysis, the
antiquark modifications will become more similar to those in EKRS.

\item The NMC data set on $F_2^A/F_2^{\rm C}$ \cite{Arneodo:1996rv},
which imposes quite stringent constraints for the $A$-systematics in
the EKRS analysis, is not used in HKM. As a result, $R_{F_2}^A$ has
less shadowing in HKM than in EKRS (see also Fig. 1 of
\cite{Eskola:2002us}), especially for heavy nuclei.

\item In addition to the recent DIS data sets, some older ones are
used in the HKM analysis. The older sets are not included in EKRS.
This, however, is presumably not causing any major differences between
EKRS and HKM, since the older data come with larger errorbars, and are
therefore typically of less statistical weight in the $\chi^2$
analysis.

\item The HKM analysis does not make use of the NMC data
\cite{Arneodo:1996ru} on the $Q^2$ dependence of $F_2^{\rm
Sn}/F_2^{\rm C}$. As explained above, these data are the main
experimental constraint for the nuclear gluons in the EKRS
analysis. Fig.~\ref{fig:snc_eks_hkm} shows a comparison between the
EKRS (solid), the HKM (dotted-dashed) results and the NMC data on
$F_2^{\rm Sn}/F_2^{\rm C}$ as a function of $Q^2$. As observed there,
the HKM results do not reproduce the measured $Q^2$ dependence of
$F_2^{\rm Sn}/F_2^{\rm C}$ at the smallest values of $x$.  This figure
demonstrates explicitly that the nuclear modifications of gluon
distributions at $0.02\lsim x\lsim 0.1$ {\em can} be pinned down with
help of these NMC data.

\begin{figure}[htb]
\begin{center}
\vspace{0cm}
\includegraphics[width=12cm]{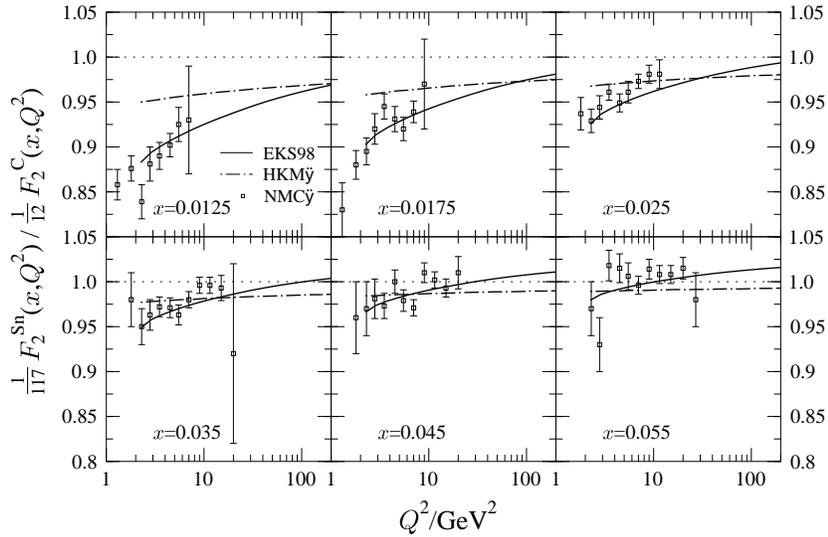} 
\vspace{-1cm}
\caption{ The ratio $F_2^{\rm Sn}/F_2^{\rm C}$ as a function of $Q^2$
at different fixed values of $x$. The data is from NMC
\cite{Arneodo:1996ru} and the curves are EKS98 \cite{Eskola:1998df}
(solid) and HKM \cite{Hirai:2001np} (dotted dashed).  }
\label{fig:snc_eks_hkm}
\end{center}
\end{figure}

\end{itemize}

\section{Future prospects}

\subsection{Constraints from $pA$ collisions at LHC, RHIC and SPS}
\label{future}
The hard probes in $pA$ collisions, especially those in LHC, will play
a major role in probing the nPDF at regions not accessed before. This
is sketched in Fig.~\ref{fig:q2_vs_x}, showing the regions in $x$ and
$Q^2$ probed by certain hard processes at different cms-energies.  Let
us consider the figure in more detail:

\textbullet~The measurements of semileptonic decays of $D$ and $\bar
D$ mesons in $pA$ will help in pinning down the nuclear gluon
distributions \cite{Lin:1995pk, Eskola:2001gt}.  Borrowed from a LO
analysis \cite{Eskola:2001gt}, the thick solid lines in
Fig.~\ref{fig:q2_vs_x} show how the average scale $Q^2 = \langle
m_T^2\rangle$ of open charm production is correlated with the average
fractional momentum $x=\langle x_2\rangle$ of the incoming nuclear
gluon in dimuon production from correlated $D\bar D$ pairs at the LHC
(computed for $p$Pb at $\sqrt s = 5500$~GeV, $2.5\le y_{\mu} \le 4.0$,
with no rapidity shifts), at RHIC ($p$Au at $\sqrt s =200$~GeV,
$1.15\le y_{\mu} \le 2.44$) and at the SPS ($p$Pb at $\sqrt s = 17.3$~GeV,
$0\le y_{\mu} \le 1$). For each solid curve, the smallest $Q^2$ shown
corresponds to a dimuon invariant mass $M=1.25$ GeV, and the the
largest $Q^2$ to $M=9.5$ GeV (LHC), 5.8 GeV (RHIC) and 4.75 GeV (SPS)
(cf. Fig. 5 in \cite{Eskola:2001gt}). We observe that data from the SPS,
RHIC and LHC will offer constraints in different regions of $x$: the
data from NA60 in the SPS will probe the antishadowing of nuclear gluons
and the RHIC data the beginning of the gluon shadowing, while the LHC
data will constrain the nuclear gluons at very small $x$, deep in the
shadowing region.

\textbullet~ Similarly, the DY cross sections of dimuons in $pA$ are
expected to set more constraints on the nuclear effects of sea quark
distributions. In $2\rightarrow 2$ kinematics $x_2=(M/\sqrt s)e^{-y}$,
where $M$ and $y$ are the invariant mass and rapidity of the lepton
pair.  The rapidity of the pair is always between the rapidities of
the leptons forming the pair. The shaded regions in
Fig.~\ref{fig:q2_vs_x} illustrate the regions in $x=x_2^A$ and $Q^2=M^2$
probed by the DY dimuons in $pA$ collisions at the LHC, RHIC and SPS.
The parameters used for the LHC are $\sqrt s=5500$~GeV, $2.5\le y\le
4.0$ with no rapidity shifts \cite{MORSCH}. Note that from the point
of view of $x_2^A$ (but {\em not} of $x_1^p$), Pb+Pb collisions at
$\sqrt s_{\rm PbPb}=5500$~GeV at the LHC are equivalent to $p$Pb collisions at
$\sqrt s_{p\rm Pb}=8800$~GeV: the decrease of $x_2$ due to the
increase in $\sqrt s$ is compensated by an increase from the rapidity
shift $y_0$ \cite{MORSCH}: $x_2^{pA} = (M/\sqrt s_{pA})
e^{-(y_{AA}+y_0)} = (M/\sqrt s_{AA})e^{-y_{AA}}=x_2^{AA}$.  For RHIC,
the shaded region corresponds to $\sqrt s = 200$~GeV, and $1.2\le y\le 2.2$,
which is the rapidity acceptance region of the forward muon arm in PHENIX 
\cite{Nagle:2002ib}.  For the SPS, we have again used $\sqrt s=17.3$~GeV and
$0\le y\le 1$. Single muon $p_T$-cuts (see
\cite{BOTJE_proc,Nagle:2002ib}) have not been applied, the shaded
regions all correspond to $M^2\ge1$~GeV$^2$.  At the highest scales shown, the
dimuon regions at the SPS and RHIC are limited by the end of the phase
space.  The SPS data may shed more light on the question of an EMC
effect for the sea quarks \cite{Eskola:2000xv}. The DY dimuons at the
LHC will probe the sea quark distributions deep in the shadowing
region and also at high scales, especially also at $Q^2=M_Z^2$.  The
region of RHIC data is again conveniently in between the SPS and LHC,
mainly probing the beginning of the shadowing region of sea quarks.
In $Ap$ collisions at the LHC, the DY dimuons with $y=(y_{\min}+y_{\rm
max})/2$ probe the regions plotted with the dashed line (the spread
due to the $y$ acceptance is not shown).

\textbullet~ The dotted line shows the kinematical region probed by
open heavy quark production in $pA$ (and also in $AA$), when both
heavy quarks are at $y=0$ in the ALICE detector frame. In this case
again, assuming $2\rightarrow 2$ kinematics, $x_2^{pA} = x_2^{AA} =
2m_T/\sqrt s_{AA}$. To illustrate the effect of moving towards forward
rapidities, also the case $y_Q=y_{\overline Q}=3$ is shown. Only the
region at $Q^2\gsim m_b^2$ (above the upper horizontal line) is probed
by $b\bar b$ production. Due to the same kinematics as in open
$Q\overline Q$ production, the dotted lines also correspond to the
regions probed by direct photon production.

\textbullet~ As demonstrated by Fig.~\ref{fig:q2_vs_x}, the hard
probes in $pA$ collisions at RHIC and at the LHC in particular will
provide us with very important constraints for the nPDF at scales
where the DGLAP evolution can be expected to be applicable. Power
corrections in the evolution \cite{Gribov:tu,Mueller:wy} can, however,
be expected to play an increasingly important role towards small
values of $x$ and $Q^2$. The effects of the first of such corrections,
the GLRMQ terms \cite{Gribov:tu,Mueller:wy} leading to nonlinear
DGLAP+GLRMQ evolution equations, have been recently studied in
\cite{Eskola:2002yc} in light of the HERA data.  From the point of
view of the DGLAP evolution of the (n)PDF, saturation of gluons takes
place when their evolution becomes dominated by the power corrections
\cite{Qiu_proc}.  Fig.~\ref{fig:q2_vs_x} shows the saturation limits
obtained for the free proton and for Pb in the DGLAP+GLRMQ approach
\cite{KOLHINEN} (solid curves; the dotted extrapolations are merely
for guiding the eye).  The saturation limit obtained for the free
proton in the DGLAP+GLRMQ analysis should be taken as an upper limit
in $Q^2$ but it is constrained quite well by the HERA data (see
\cite{Eskola:2002yc}).  In obtaining the saturation limit shown for
the Pb nucleus in \cite{KOLHINEN}, the constraints from the $Q^2$
dependence of $F_2^{\rm Sn}/F_2^{\rm C}$, however, have not yet been
taken into account. For a comparison of the saturation limits obtained
in other models, see \cite{KOLHINEN}.

\textbullet~ Fig.~\ref{fig:q2_vs_x} shows also which hard probes in
$pA$ collisions can be expected to probe the nPDF in the gluon
saturation region. Especially the probes directly sensitive to the
gluon distributions are interesting from this point of view. Such
probes would be open $c\bar c$ production at small $p_T$, and direct
photon production at $p_T\sim $ few GeV, both at as forward rapidities
as possible (see the dotted lines). Note that open $b\bar b$
production is already in the applicability region of the linear DGLAP
evolution. In light of the saturation limits shown, the chances for
measuring the effects of nonlinearities in the evolution through open
$c\bar c$ production in $pA$ at RHIC would seem marginal.  At the LHC,
however, measuring saturation effects in the nuclear gluon
distributions through open $c\bar c$ in $pA$ could be possible.

\subsection{Improvements of the DGLAP analyses}
On the practical side, the global DGLAP fit analyses of the nPDF
discussed above can be improved in obvious ways. The EKRS analysis
should be made more automatic and a proper statistical treatment, such
as in HKM, should be added.  The automatization alone is, however, not
expected to change the nuclear modifications of the PDF significantly
from EKS98 but more quantitative estimates of the uncertainties and of
their propagation would be obtained.  This work is in progress. As
also discussed above, more data contraints should be added to the HKM
analysis.  More generally, all presently available data from hard
processes in DIS and $pA$ collisions have not yet been exhausted: for
instance, the recent DIS data for $F_2^{\nu{\rm Fe}}$ and
$F_3^{\nu{\rm Fe}}$ from $\nu$Fe and $\bar \nu$Fe collisions by
CCFR\cite{Seligman:mc} (not used in EKRS or in HKM), could help in
pinning down the valence quark modifications \cite{Botje:1999dj}. In
the future, the hard probes in $pA$ collisions at the LHC, RHIC and
the SPS will offer very important constraints for the nPDF, especially for
gluons and sea quarks. Eventually, the nPDF DGLAP analyses should be
extended to NLO perturbative QCD. As discussed above, the effects of
power corrections \cite{Gribov:tu,Mueller:wy} to the DGLAP equations
and also to the cross sections \cite{Guo:2001tz}
should be analysed in detail in context with the global fits to the
nuclear data.

\vspace{1cm}
\noindent{\large \bf Acknowledgements.}  We thank N. Armesto, F. Gelis,
P.V. Ruuskanen, I. Vitev and other participants of the CERN Hard
Probes workshops for discussions. We are grateful for the Academy of
Finland, Project 50338, for financial support.  C.A.S. is supported by
a Marie Curie Fellowship of the European Community programme TMR
(Training and Mobility of Researchers), under the contract number
HPMF-CT-2000-01025.

\end{document}